# Thermodynamics and Relativity:
# A Revised Interpretation of the Concepts of Reversibility and Irreversibility

Jean-Louis Tane (tanejl@aol.com)
Formerly with the Department of Geology, University Joseph Fourier, Grenoble, France

**Abstract**: It is generally admitted in thermodynamics that, for a given change in volume, the work done by a system is greater in conditions of reversibility than in conditions of irreversibility. If the basic conventions of physics are strictly respected, it seems that this proposition needs to be reversed. Beyond this formal aspect, the discussion consolidates the idea that thermodynamics and relativity are closely connected.

## - 1 - Statement of the problem and brief reminder of the classical interpretation

Let us imagine a thermodynamic system defined as a given amount of gas contained in a cylindrical vessel. We suppose that the vessel is open upwards, but that the gas is separated from the surroundings by a mobile frictionless piston, of negligible weight.

If the external pressure $P_e$ (pressure of the surroundings) equals the internal pressure $P_i$ (pressure of the gas), the piston remains motionless. If they are different, the piston moves until they become equal, so that the volume of the gas increases or decreases by a quantity $\Delta V$. The condition $P_e > P_i$ leads evidently to $\Delta V < 0$ and the condition $P_i > P_e$ to $\Delta V > 0$.

According to the basic conventions of physics, the equation linking the change in work of the system to its change in volume takes the general form:

$$dW = - P_e \, dV \qquad (1)$$

whose precise meaning is therefore:

$$dW_{syst} = - P_{surr} \, dV_{syst}$$

As concerns equation 1, two preliminary remarks can be made:

a) The sign minus is preferred to the sign plus in order to express the fact that the system receives work from the surroundings ($dW > 0$) when its volume decreases and does work on the surroundings ($dW < 0$) when its volume increases.

b) The pressure taken into account is $P_e$, not $P_i$, because $P_e$ corresponds to the real case of an irreversible process. It is exclusively in the limited case of a reversible process (i.e. when $P_e$ is constantly equivalent to $P_i$) that equation 1 can be written under the form:

$$dW = - P_i \, dV \qquad (2)$$

The difference between a reversible and an irreversible process is analyzed with more or less details in most books of Physical Chemistry. Among the best known is the treatise by Atkins and de Paula [1] from which Fig.1 is inspired.

This figure provides a schematization of the interpretation usually admitted for the isothermal expansion of a perfect gas. The upper curve corresponds to a reversible process, for which $P_e = P_i = nRT/V$, the lower curve to an irreversible process for which $P_e$ represents the atmospheric pressure, which can be considered constant during the time of the experiment. The gas being in expansion ($\Delta V > 0$), we easily conceive that the average value of $P_i$ is greater than the constant value of $P_e$. The symbols $V_i$ and $V_f$, corresponding respectively to the points M and N, designate the initial and final volume of the gas.

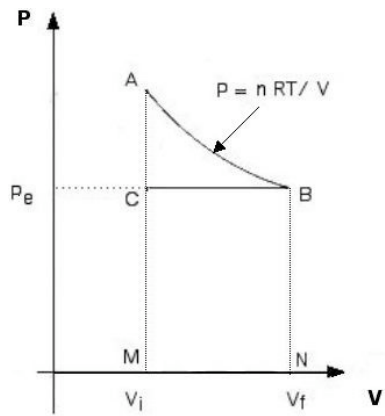

Fig. 1

Observing that the area under the curve $P_i$ (area ABNM) is greater than the area under the line $P_e$ (area CBNM), the conclusion classically deduced is that, for a given change in volume, the work done in conditions of reversibility is greater than the work done in conditions of irreversibility.

As explained below, this conclusion is debatable, because an alternative interpretation is possible.

## 2. Alternative interpretation

An important point of the thermodynamic reasoning is that the entirety "system + surroundings" is itself implicitly considered as an isolated system. This is a way to say that the exchanges between the system and the outside are supposed limited in space, rather than extended to the whole universe.

In most cases, the limits of the surroundings are not known, so that their extensive parameters (mass or volume) are not known either and only their intensive parameters, such as pressure and temperature, are taken in consideration. The fact that they are often interpreted as constants means that the surroundings are supposed much greater than the system (situation realized when the surroundings correspond to the atmosphere) or that a mechanism maintains their intensive parameters at constant values.

As will be seen now, interesting information can obtained when our knowledge of the parameters reaches the same level of precision for the system and its surroundings. Let us imagine than instead of being open upwards, the cylindrical vessel evoked above is totally isolated, and composed of two parts called 1 and 2. We suppose that part 1 contains a gas whose pressure is $P_1$, part 2 a gas whose pressure is $P_2$ and that both parts are separated by a piston initially locked.

If the piston is liberated, it will move towards the part of lower pressure until the pressures become equal. In the thermodynamic analysis of the process, we can consider that part 1 represents the system and part 2 the surroundings, or that part 2 represents the system and part 1 the surroundings. The important point to keep in mind is that the whole system, which corresponds to the set "part 1 + part 2", has been defined as isolated.

Applying equation 1 to part 1, we get

$$dW_1 = -P_2 dV_1 \qquad (3)$$

If the total volume change of part 1 is $\Delta V_1$, the corresponding change in work is:

$$W_1 = \int_{V_1}^{V_2} -P_2 dV_1 \qquad (4)$$

In the context presently examined, we easily conceive that $P_2$ varies during the process. Nevertheless, designating its average value by the symbol $P_2^*$, we can write:

$$W_1 = -P_2^* \Delta V_1 \qquad (5)$$

In eq. 5, $W_1$ represents the change in work of part 1. According to the convention in use, the notation $W_1$ is preferred to the notation $\Delta W_1$, in order to distinguish the parameters which are considered as state functions (such as V) from those which are not (such as W). Being the measure of a pressure, $P_2^*$ is necessarily positive, so that $W_1$ is positive when $\Delta V_1$ is negative and conversely.

Similarly, the change in work of part 2 is

$$W_2 = -P_1^* \Delta V_2 \qquad (6)$$

Although the numerical values of $P_1^*$ and $P_2^*$ are generally unknown, there is evidence that a positive value of $\Delta V_1$ implies the condition $P_1^* > P_2^*$ and a negative value of $\Delta V_1$ the condition $P_1^* < P_2^*$.

At this stage of the discussion, two important propositions can be advanced:

**First information**: *the change in work of the system is greater in conditions of irreversibility than it would be in conditions of reversibility*.

Indeed, if the system is defined as being part 1, we can write:

$$W_{1irr} = -P_2^* \Delta V_1 \qquad (7)$$

$$W_{1rev} = -P_1^* \Delta V_1 \qquad (8)$$

As a result:

$$W_{1irr} - W_{1rev} = (P_1^* - P_2^*) \Delta V_1 \qquad (9)$$

Since $\Delta V_1$ is positive when $P_1^* > P_2^*$ and negative when $P_1^* < P_2^*$, the quantity $W_{1irr} - W_{1rev}$ is always positive so that, in all cases, we have:

$$W_{1irr} > W_{1rev} \tag{10}$$

If the system is defined as being part 2, applying the same reasoning to $W_{2irr.}$ and $W_{2rev}$, leads to:

$$W_{2irr} > W_{2rev} \tag{11}$$

**Second information**: *although the whole system is isolated, it appears that its total balance in work is positive when the internal exchange of work is irreversible, and reaches a zero value exclusively when this exchange is reversible.*

The whole system being defined as the set "part 1 + part 2", its total balance in energy which can be designated $W_{syst}$, is:

$$W_{syst} = W_1 + W_2 \tag{12}$$

If the exchange of work between part 1 and part 2 is irreversible, the global energy change of the system is:

$$W_{syst.irr} = W_{1irr} + W_{2irr} = - P_2^* \Delta V_1 - P_1^* \Delta V_2$$

Since $\Delta V_2 = - \Delta V_1$, this result is equivalent to:

$$W_{syst.irr} = W_{1irr} + W_{2irr} = \Delta V_1 (P_1^* - P_2^*) \tag{13}$$

Having already noted that $\Delta V_1$ is positive when $P_1^* > P_2^*$ and negative when $P_1^* < P_2^*$, the final result can always be written:

$$W_{syst.irr} > 0 \tag{14}$$

In a similar way, if the exchange of work between part 1 and part 2 is reversible, the global energy change of the system, noted $W_{syst.rev}$, is:

$$W_{syst.rev} = W_{1rev} + W_{2rev} = \Delta V_1 (P_2^* - P_1^*) \tag{15}$$

Remembering that the condition of reversibility implies $P_2^* = P_1^*$, we are led to the conclusion that:

$$W_{syst.rev} = 0 \tag{16}$$

## 3. Graphical representation of these considerations

The graphical representation previously examined (Fig.1) deals with the isothermal expansion of a perfect gas against the atmosphere. Having in mind this context, we can give the gas the designation "part 1", and the fraction of the atmosphere which exchanges energy with the gas the designation "part 2".

Although many characteristics of part 2 are unknown, the whole system defined as "part 1 + part 2" is isolated, so that the exchanges in energy occurring inside it appear as a particular case of the general situation discussed in section 2.

The peculiarities lie in the fact that the gas present in part 1 is perfect and that $P_2$ can be considered constant, being the pressure of the atmosphere. This second peculiarity infers that $P_2^* = P_2$ and that the work exchanged between part 1 and part 2 is greater than it would be when $P_2$ varies, because this variation is a decrease.

Similarly, $T_2$ can be considered constant, being the temperature of the atmosphere, so that $T_2^* = T_2$. Depending on whether $T_2$ is constant or variable, the heat exchanged between part 1 and part 2 is not the same. This last condition infers itself that the work exchanged between part 1 and part 2 is not the same either. Despite these reservations, whose incidence on thermodynamic calculations are well known, the important point of the discussion is that the equations considered in section 2 are valid in all cases, since they correspond to a general situation.

The reason why the conclusion classically derived from Fig.1 is questionable comes from the fact that this figure is not a representation of the quantity

$$W_1 = \int_{V_1}^{V_2} - P_2 dV_1 \qquad (4)$$

it is a representation of the quantity

$$W_1 = \int_{V_1}^{V_2} P_2 dV_1$$

As already seen, eq. 4 leads to eq.7 and 8, whose expressions are $W_{1irr} = - P_2^* \Delta V_1$ and $W_{1rev} = - P_1^* \Delta V_1$. Their graphical representations are given in Fig. 2 where (referring to the expansion of the gas present in part 1), the area C'B'NM corresponds to the work $W_{1irr}$ (irreversible process) and the area A'B'NM to the work $W_{1rev}$ (reversible process). The fact that the first area is smaller than the second means that $W_{1irr}$ is less negative than $W_{1rev}$, a proposition which can be written $W_{1irr} > W_{1rev}$ and therefore is in accordance with eq. 10.

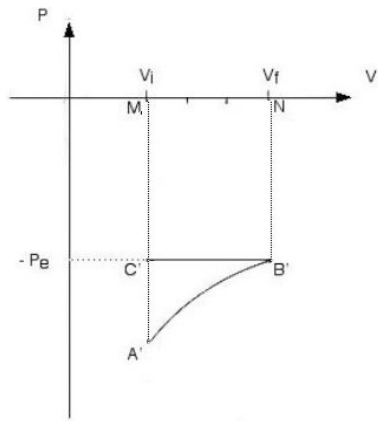

Fig. 2

Concerning $W_2$, the starting equations are $W_{2irr} = - P_1^* \Delta V_2$ and $W_{2rev} = - P_2^* \Delta V_2$. Remembering that $\Delta V_2 = - \Delta V_1$, they can be written $W_{2irr} = P_1^* \Delta V_1$ and $W_{2rev} = P_2^* \Delta V_1$. This is illustrated in Fig 3, which has some similarity with Fig.1, except that the external pressure $P_e$ to take into account is not the atmospheric pressure (as suggested by Fig 1), but the average pressure $P_1^*$. Having $P_1^* > P_2^*$, we get $W_{2irr} > W_{2rev}$, a result already noted in equation 11 and whose graphical representation corresponds to the fact that the area ADNM is greater than the area CBNM.

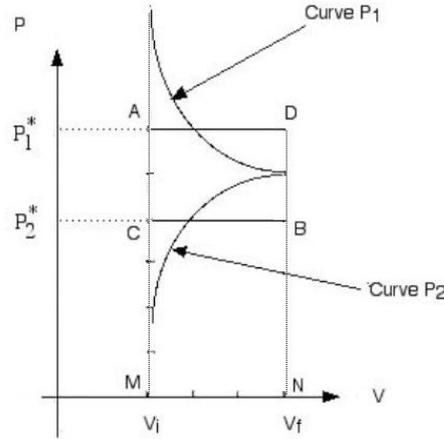

Fig. 3

For an easier checking of these considerations, let us imagine that inside the isolated system examined in section 2, the value of $\Delta V_1$ (expansion of the gas present in part 1) is 3 units of volume, the value of $P_1^*$ is 6 units of pressure and the value of $P_2^*$ is 4 units of pressure. This situation corresponds to the one schematically drawn on Fig. 3 and leads to the followings results:

For part 1, we have :  $W_{1irr} = - P_2^* \Delta V_1$ (eq.7),  i.e. $W_{1irr} = - 4 \times 3 = -12$
$W_{1rev} = - P_1^* \Delta V_1$ (eq.8),  i.e. $W_{1rev} = - 6 \times 3 = -18$
so that  $W_{1irr} - W_{1rev} = -12 - (-18) = 6$

For part 2, we have :  $W_{2irr} = - P_1^* \Delta V_2 = P_1^* \Delta V_1 = 6 \times 3 = 18$
$W_{2rev} = - P_2^* \Delta V_2 = P_2^* \Delta V_1 = 4 \times 3 = 12$
so that  $W_{1irr} - W_{1rev} = 18 - 12 = 6$

In both cases, we see that the irreversible work is greater than the reversible work, a conclusion which is an illustration of the "**first information**" mentioned in section 2, when the system taken in consideration is successively "part 1" and "part 2".

Now, if we consider the whole system defined as the set "part 1 + part 2", the obtained results are the following:

For an irreversible process, we know from eq. 12 and 13 that:

$$W_{syst.irr} = W_{1irr} + W_{2irr} = \Delta V_1 (P_1^* - P_2^*)$$
so that we get  $W_{syst.irr} = 3 (6 - 4) = 6$

Being positive, this result is an illustration of the **"second information"** mentioned in section 2

For a reversible process, we know from eq. 12 and 15 that:

$$W_{syst.rev} = W_{1rev} + W_{2rev} = \Delta V_1 (P_2^* - P_1^*)$$

Since reversibility is a particular case of irreversibility, the formula is evidently the same as the previous one but, remembering that the conditions of reversibility imply the equality $P_2^* = P_1^*$, we get for the final result:

$$W_{syst.rev} = 0$$

## 4. Suggested improvement in the wording of the laws of thermodynamics

The alternative interpretation presented in sections 2 and 3 introduces some modifications of our usual conception of the thermodynamic theory.

In conventional thermodynamics, an isolated system is characterized by the fact that its global change in energy is zero, whatever the processes occurring inside it are reversible or irreversible.

Since the whole system considered above was defined as isolated, we can be tempted to think that the positive value of $W_{syst.irr}$ (eq.14) is necessarily compensated by a negative value of $Q_{syst.irr}$, which represents the sum of the exchanges of heat between part 1 and part 2. Discussions about this problem are presented in all courses of thermodynamics, for instance in the well-known book by E. B. Smith [2] from which the comment evoked below is borrowed. Referring to a system which is not isolated but closed (it is the case for the context examined above if the system is defined as being part 1, and the surroundings part 2), the reasoning reminded by the author can be summarized as follows.

Admitting that the gas contained in part 1 is in expansion (a situation which implies the conditions $\Delta V_1 > 0$ and $P_1^* > P_2^*$), the attention is called to the fact that, for a given change of state of the system, both $W_{1irr}$ and $W_{1rev}$ are negative, but $W_{1irr}$ is less negative that $W_{1rev}$. The conclusion is therefore written $W_{1rev} < W_{1irr}$, a relation which is in perfect accordance with eq. 10 and Fig. 2 of the present paper.

In the book by Smith, the discussion is presented at the scale of an elementary change in energy, so that the relation $W_{1rev} < W_{1irr}$ takes the form $dW_{1rev} < dW_{1irr}$. Then referring to the usual understanding of the first law of thermodynamics which postulates the equality

$$dU_{1rev} = dU_{1irr}, \qquad (17)$$

the author concludes that the condition $dW_{1rev} < dW_{1irr}$ is necessarily compensated by the condition

$$dQ_{1rev} > dQ_{1irr} \qquad (18)$$

Although this conclusion appears evident, it is not sure that it is perfectly consistent with the second law of thermodynamics. The reason is the following.

The second law is classically presented through the expression:

$$dS = dQ/T + dS_i \qquad (19)$$

whose precise meaning is:

$$dS = dQ/T_e + dS_i \qquad (20)$$

A well-known peculiarity of these equations is that $dS_i$ has a positive value in the case of an irreversible process and a zero value in the case of a reversible process.

Eq. 19 has the dimension of an entropy, but takes the dimension of an energy if it is written under the form:

$$T_e dS = dQ + T_e dS_i \qquad (21)$$

In eq. 21, the term $T_e dS_i$ is necessarily positive since $T_e$ is an absolute temperature and $dS_i$ is positive, as already noted. As a consequence, this equation implies the relation:

$$T_e dS > dQ \qquad (22)$$

Comparing eq. 18 with eq. 22, the idea which comes spontaneously in mind is that the terms $dQ_{1rev}$ and $dQ_{1irr}$ of eq. 18 correspond respectively to the terms $T_e dS$ and $dQ$ of eq. 22.

Although this interpretation is the one classically admitted in thermodynamics, it is totally dependant on our belief in relation 17, which is itself a postulate expressing the usual understanding of the first law. Obviously, the convention adopted for the change in work and the change in heat is not the same: in the first case, the term $-P_e dV$ is assigned to an irreversible process, in the second case, the term $T_e dS$ is assigned to a reversible process. From the logical point of view, this change of convention is not totally satisfactory. The situation is similar to the one evoked earlier concerning the use of Fig. 1 instead of Fig. 2.

An alternative interpretation has been advanced recently [3]. It suggests that the global energy change of a system, noted $dU_{syst}$, be written in the extended form:

$$dU_{syst} = dU_e + dU_i \qquad (23)$$

where $dU_e$ represents the energy exchanged between the system and its surroundings (external energy) and $dU_i$ the energy created, or destroyed, inside the system (internal energy).

A detailed explanation of eq. 23 can be found in ref ([3], [4]) and in the previous papers quoted in them. The main points to keep in mind are the following:

- this equation represents a synthesis of the first and second laws of thermodynamics. It implies a positive value of $dU_i$ (at least for the systems exclusively made of inert matter).

- the term $dU_i$ is assimilated to the differential of the Einstein mass-energy relation, ($E = mc^2$), a proposition which can be summarized through the equation:

$$dU_i = dE = -c^2 dm \qquad (24)$$

- in eq. 24, the sign minus is preferred to the sign plus, because it corresponds to the idea that a disintegration of mass leads to a creation of energy and conversely. The situation is comparable with the one encountered above concerning the sign minus of eq. 1.

- in conventional thermodynamics, the term $dU_i$ of eq. 23 is not taken into account, so that no distinction is made between $dU_{syst}$ and $dU_e$. They are designated by the same symbol $dU$ and called "the change in internal energy".

- referring to eq 23, an isolated system corresponds to the condition $dU_e = 0$ (no exchange between the system and its surroundings) but we have $dU_i \geq 0$. When the exchanges of energy occurring within the system are reversible the condition is $dU_i = 0$, when they are irreversible the condition is $dU_i > 0$.

- By integration, eq. 23 leads to:

$$\Delta U_{syst} = \Delta U_e + \Delta U_i \qquad (25)$$

and eq. 21 to:

$$T_e^* \Delta S = \Delta Q + T_e^* \Delta S_i \qquad (26)$$

where $T_e^*$ is the average value of $T_e$ during the process taken in consideration.

- Eq. 26 represents the form taken by eq. 25 in the case of an exchange of heat. Each term of the second equation has the same meaning as the corresponding term of the first equation. Taking into account the comment made above about the parallelism between eq. 18 and 22, eq. 26 can be understood as having the significance:

$$Q_{irr} = Q_{rev} + \Delta Q_i \qquad (27)$$

In a similar way, eq. 25 has the significance:

$$\Delta U_{irr} = \Delta U_{rev} + \Delta U_i$$

- Concerning an exchange of work, the peculiarity to take into account is that eq. 9 can be given the form:

$$W_{1irr} = W_{1rev} + (P_1^* - P_2^*) \Delta V_1 \qquad (28)$$

and more generally the form:

$$W_{irr} = W_{rev} + \Delta W_i \qquad (29)$$

Remembering that the term $(P_1^* - P_2^*) \Delta V_1$ is always positive, $\Delta W_i$ is always positive too, so that eq. 29 appears as the form taken by eq. 25, in the case of an exchange of work.

**NB**: A numerical example of heat exchange is presented in ref [4] showing the difference between the classical interpretation of the process (integration of eq. 21) and the new suggested one (integration of eq. 23).

**5. Conclusions**

It is important to emphasize that the interpretation presented above is not a rejection of the laws of thermodynamics, but an extension of their significance.

The novelty introduced is mainly theoretical. It contributes to a better consistency between the first and the second law and confirms the idea already advanced [3] that thermodynamics and relativity are closely connected.

**References**

[1]  P. W. Atkins and J. de Paula. *Physical Chemistry*, Oxford University Press, 8th Edition, 2006 (p. 36)

[2]  E. B. Smith. *Basic Chemical Thermodynamics*. Imperial College Press, 5$^{th}$ Edition, 2004 (p. 21)

[3]  J-L Tane. *Thermodynamics and Relativity: A Condensed Explanation of their Close Link*. arxiv.org/pdf/physics/0503106, March 2005

[4]  J-L Tane. *Thermodynamics and Relativity: a Short Explanation of their Close Link* The general Science Journal, April 2007 (summarized version of the previous paper)